\documentclass{jps-cp}

\def\be{\begin{equation}}
\def\ee{\end{equation}}

\title{Kondo Temperature and High to Low Temperature Crossover in Impurity Models of Correlated Electrons}

\author{V\'aclav  \textsc{Jani\v s} and Anton\'\i n \textsc{Kl\'\i\v c}}

\inst{Institute of Physics, The Czech Academy of Sciences, Na Slovance 2, CZ-18221 Praha  8,  Czech Republic}

\email{janis@fzu.cz}

\recdate{September 2, 2019}

\abst{Kondo temperature is standardly defined from the local zero-temperature susceptibility in the regime of strong electron correlations as a new scale controlling the low-temperature asymptotics of thermodynamic quantities of diluted impurities. We show by using a two-particle self-consistent theory that the Kondo temperature can be identified as a crossover temperature at which the zero-temperature quantum fluctuations equal the thermal ones in the electron-hole correlation function. The high-temperature Curie-Weiss susceptibility is shown to go over to the Pauli one below the Kondo temperature.}

\kword{Interacting impurity models, two-particle self-consistency, Kondo temperature, quantum and thermal fluctuations}

\begin{document}
\maketitle

\section{Introduction}

Kondo temperature was first determined from the logarithmic divergence of the low-temperature limit of the perturbation expansion in the $s-d$ model \cite{Kondo:1964aa}. There is no divergence in the exact solution, and the Kondo temperature is defined more consistently from the zero-temperature susceptibility of the strong-coupling limit of the single-impurity $s-d$ \cite{Andrei:1983aa} or Anderson models \cite{Tsvelick:1983aa}. Presently, the Kondo temperature generally determines an energy scale controlling the low-temperature asymptotics of thermodynamic and spectral functions of diluted magnetic systems with strong electron correlations. A question remains whether we can give the Kondo temperature a meaning of a crossover point between high- and low-temperature regimes. We show that the Kondo temperature can indeed be defined as a crossover point in the two-particle vertex where thermal and quantum fluctuations are equal.

 The Kondo scale related to the Kondo temperature is generically a two-particle quantity controlling the proximity to a singularity in the electron-hole correlation function \cite{Janis:2007aa}. To suppress crossing the singularity of the Hartree approximation and to reach the Kondo regime of impurity models, one has to use a better self-consistent solution.    Attempts to suppress the singularity at finite interaction strength by one-particle self-consistency with a dynamical self-energy in approximations of fluctuation-exchange type failed to reproduce the Kondo asymptotics \cite{Hamann:1969aa}. Hence, the only consistent way to describe reliably the strong-coupling limit is to introduce nontrivial renormalizations of vertices. The best and most straightforward way to introduce renormalizations of vertex functions and two-particle self-consistency is the parquet construction \cite{DeDominicis:1964aa,Bickers:1991aa}. This is, however, a tremendous task if we want to keep active all the  degrees of freedom of the vertex functions. Further approximations and modifications of the parquet equations are needed to make them usable for the description of the Kondo regime in impurity models.

 A natural simplification emerges at the critical region of the divergence of the vertex function where only small transfer frequencies remain relevant. Next, we can neglect the fluctuations    in fermionic variables, irrelevant at zero temperature. The parquet equations can then be reduced to a static, mean-field-like approximation for the irreducible vertex from the electron-hole channel \cite{Janis:2007aa}. We recently improved upon this simple scheme in that we consistently matched the two-particle and one-particle thermodynamic and spectral quantities \cite{Janis:2017aa,Janis:2017ab}. Finally, we extended this approximation to non-zero temperatures and thus opened a way for studying the genesis of the Kondo effect and the crossover form the classical, high-temperature state to the quantum, zero-temperature one \cite{Janis:2019aa}. Similar static approximations on two-particle vertices were used in Refs.~\cite{Vilk:1997aa,Kusunose:2010aa}. They both separately approximate charge and spin fluctuations by effective interactions in response functions and reach a two-particle self-consistency either by obeying local sum rules, Ref.~\cite{Vilk:1997aa}, or the crossing symmetry of vertex functions, Ref.~\cite{Kusunose:2010aa}. Neither of these approaches is, however, able to deliver the Kondo scale analytically. Our scheme does that, since it is simpler. It approximates only a single irreducible vertex and utilizes a simplifying low-energy scaling in the critical region of the divergence in one of the Bethe-Salpeter equations.


\section{Effective-Interaction Approximation}

\begin{figure}[tbh]
\begin{minipage}{0.55\textwidth}\includegraphics[width=8.5cm]{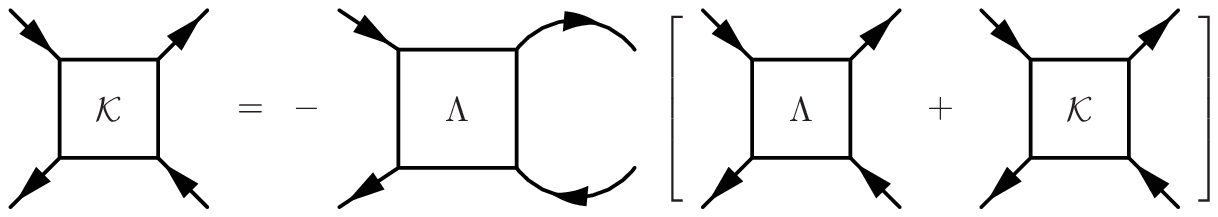}\end{minipage}\hspace{4pt},\hspace{4pt}\begin{minipage}{0.4\textwidth}\includegraphics[width=6.5cm]{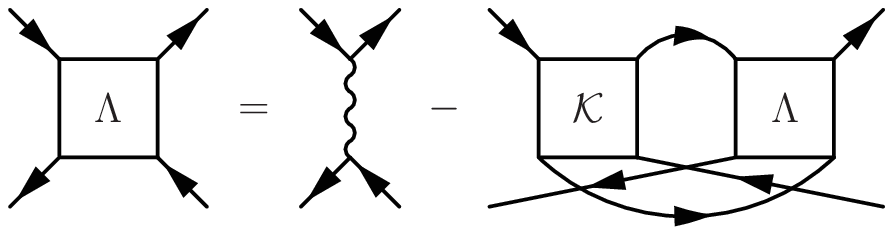}\end{minipage}
\caption{Diagrammatic representation of the reduced parquet equations with the irreducible $\Lambda$ and reducible $\mathcal{K}$ vertex from the electron-hole scattering channel. The reduced Bethe-Salpeter equation in the electron-hole channel is left and in the electron-electron channel is right. The vertical wavy line stands for the bare interaction. }
\label{fig:RPE}
\end{figure}

The approximation of Ref.~\cite{Janis:2019aa} reduces the two-channel parquet equations to a couple  of equations for the singlet irreducible $\Lambda_{\uparrow\downarrow}$  and reducible $\mathcal{K}_{\uparrow\downarrow}$ vertices from the electron-hole Bethe-Salpeter equation, see Fig.~\ref{fig:RPE} for their diagrammatic representation. 
A static approximation of the irreducible vertex $\Lambda_{\uparrow\downarrow}$ leads to a mean-field-like (static and local) theory where the bare Hubbard interaction $U$ of lattice or impurity models is replaced by an effective one, $\Lambda_{\uparrow\downarrow}$, in the Hartree solution for the thermodynamic functions and in RPA for the spectral ones. 

The effective interaction calculated from the reduced parquet equations  can be represented as follows \cite{Janis:2019aa}
\begin{equation}\label{eq:Lambda-static}
\Lambda_{\uparrow\downarrow}  =  \frac U{1 + K_{\uparrow\downarrow}X_{\uparrow\downarrow}} \,,
\end{equation}
where
\begin{equation}\label{eq:K-mp}
K_{\uparrow\downarrow} = - \Lambda_{\uparrow\downarrow}^{2}\left\langle \Im \left[G_{\downarrow}(x_{+}) G_{\uparrow}(x_{+})\right]\right\rangle_{x} \,,
\end{equation}
is the weight of the pole of the electron-hole vertex projected onto the values of the fermionic variables at the Fermi energy. We abbreviated real frequencies approached from the complex plane  $x_{\pm} = x \pm i0^{+}$ and used the angular brackets to denote a spectral integral with the Fermi distribution $f(x) = 1/(e^{\beta x} + 1)$
\begin{equation}
\left\langle G_{s}(x + \omega) G_{s'}(x + \omega')\right\rangle_{x} = -\int_{-\infty}^{\infty} \frac{dx}{\pi} f(x)  G_{s}(x + \omega) G_{s'}(x + \omega') \,.
\end{equation}
The other term in the denominator on the right-hand side of Eq.~\eqref{eq:Lambda-static},
\begin{equation}
 X_{\uparrow\downarrow}  = \int_{-\infty}^{\infty}\frac{dx}{\pi} \left\{\frac{\Re\left[ G_{\uparrow}(x_{+})G_{\downarrow}(-x_{+})\right]}{\sinh(\beta x)}  \Im\left[ \frac 1{\mathcal{D}_{\uparrow\downarrow}(-x_{+})}\right] - f(x) \Im\left[\frac {G_{\uparrow}(x_{+}) G_{\downarrow}(-x_{+})}{\mathcal{D}_{\uparrow\downarrow}(-x_{+})} \right] \right\} \,,
\end{equation}
is the leading contribution from the pole of the electron-hole vertex to screening of the bare interaction due to multiple scatterings of pairs of electrons. We used an equality $f(x) + b(x) = 1/\sinh(\beta x)$. We straightforwardly used analytic continuations of summations over Matsubara frequencies to spectral integrals with Fermi, $f(x)$, and Bose, $b(x) = 1/(e^{\beta x} - 1)$, distributions.  

The pole emerges when the positive static value ($\Omega =0$) of function 
\begin{equation}\label{eq:Det-Omega}
\mathcal{D}_{\uparrow\downarrow}(\Omega_{+}) = 1 + \Lambda_{\uparrow\downarrow} \left[\left\langle G_{\downarrow}(x + \Omega_{+}) \Im G_{\uparrow}(x_{+})\right\rangle_{x} + \left\langle G_{\uparrow}(x - \Omega_{+}) \Im G_{\downarrow}(x_{+})\right\rangle_{x}\right]   \,.
\end{equation}
reaches zero.  Due to the two-particle self-consistency it cannot happen in the impurity models at any temperature and finite interaction strength. Increasing interaction strength at the spin and charge symmetric situation ($G_{\uparrow}(x_{+}) = G_{\downarrow}(x_{+})$ and $G(x_{+}) = - G(-x_{+})$) drives the solution at very low temperatures towards a quantum critical point and exponentially small dimensionless Kondo scale $a = 1 + 2g\Lambda_{\uparrow\downarrow}$, where 
\begin{equation}\label{eq:g-full}
g = \left\langle \Im\left[G(x_{+})^{2}\right]\right\rangle_{x} =  \int_{0}^{\infty} \frac{dx}{\pi}  \tanh\left( \frac{\beta x}2\right)\Im\left[G(x_{+})^{2} \right]  < 0\,.
\end{equation}

We solve the self-consistent implicit equation for the effective interaction, Eq.~\eqref{eq:Lambda-static}, in the Kondo limit, that is for the spin and charge symmetric state with strong electron repulsion. All quantities become functions of the Kondo scale $a$. We decompose the $X$ integral into $X(a) = X_{0}(a) + \Delta X(a)$ to separate quantum and thermal fluctuations. Since the Kondo scale $a\ll 1$ we can approximate the denominator in the $X$ integrals by its leading small-frequency term $D(\omega_{+}) = a - i\omega D'(a)$. The contribution from the quantum fluctuations surviving to zero temperature in this polar approximation is %
\begin{subequations}\label{eq:X}
\begin{equation}\label{eq:X0}
    X_{0}(a)=  - \int_{0}^{\infty} \frac{dx}{\pi} \tanh\left( \frac{\beta x}2\right) \frac{\Re\left[[G(x_{+})^{2}\right]D'(a)x  +  \Im \left[G(x_{+})^{2}\right] a }{ a^{2} + D'(a)^{2}x^{2}} \,.
\end{equation}
The thermal fluctuations contribute to the effective interaction via the following integral
\be\label{eq:DeltaX}
\Delta X(a) = -\frac{2 D'(a)}\pi \int_{0}^{\infty} \frac {dx\ x}{\sinh(\beta x)}\frac{\Re\left[ G(x_{+})^{2}\right]}{a^{2} + D'(a)^{2}x^{2}}  
\ee
\end{subequations}
that linearly vanishes in the limit of low temperatures $T\to 0$. Finally, the dynamical coefficient in the expansion of the denominator, $D'(a)$, in the $X$ integrals is   
\begin{equation}\label{eq:Dprime}
D'(a) =\frac{\beta \Lambda(a)}{2\pi\Delta}\int_{0}^{\infty}\frac{dx }{\cosh^{2}(\beta x/2)} |\Im G(x_{+})|^{2} \,.
\end{equation}
Equation~\eqref{eq:Lambda-static} for the effective interaction can further be rearranged to a more suitable form 
\begin{align}\label{eq:Lambda-m}
\Lambda(a) &= \frac 1{2(1 - a) [X_{0}(a) + \Delta X(a)]}\left[\sqrt{1 + 4(1 - a)U [X_{0}(a) + \Delta X(a)]}  - 1\right] \,.
\end{align}

Integrals $X_{0}(a), \Delta X(a)$, and $D'(a)$ are functions of  parameter $a$. We close the set of he defining equations with another determining the dimensionless Kondo scale in thermal equilibrium at a given temperature $T = 1/k_{B}\beta$ and the bare interaction strength $U$  \begin{align} \label{eq:a-def}
a = 1 + g\Lambda(a)  \,.
\end{align}
We chose the unperturbed propagator $G(\omega_{\pm}) = 1/(\omega \pm i\Delta)$ for the single-impurity Anderson model, with $\Delta$ set as the energy unit, to demonstrate  the temperature behavior explicitly. 

\begin{figure}[tbh]
\begin{minipage}{0.5\textwidth}\includegraphics[width=7.5cm]{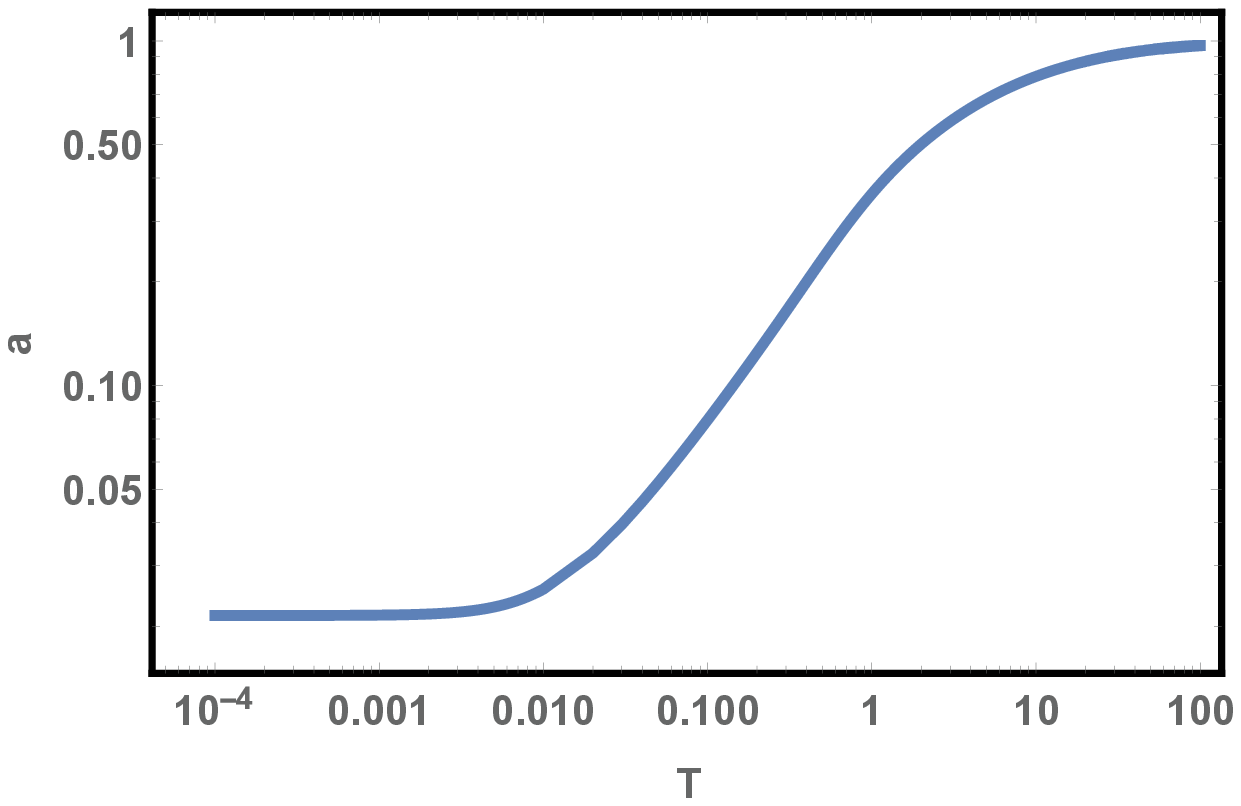}\end{minipage}\begin{minipage}{0.5\textwidth}\includegraphics[width=7.5cm]{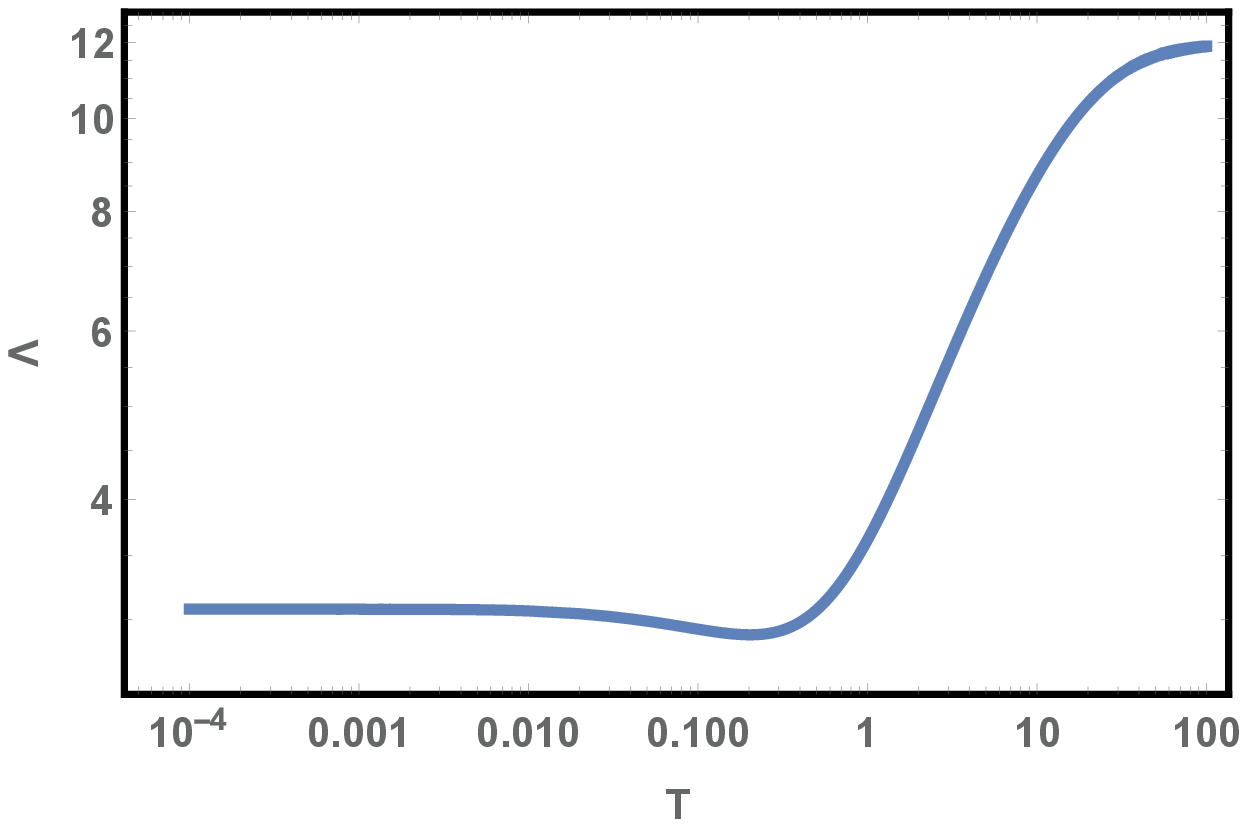}\end{minipage}
\caption{Temperature dependence of the Kondo scale $a$ (left panel) and of the effective interaction $\Lambda$ for $U=12\Delta$  in the logarithmic scale. The Kondo scale grows with temperature from its minimum at zero temperature $a_{0}>0$  to its maximum $a_{\infty}=1$ at $T=\infty$. The effective interaction approaches the bare one in the limit $T\to\infty$.}
\label{fig:A-Lambda}
\end{figure}

We use Mathematica to solve equations~\eqref{eq:X}-\eqref{eq:a-def}.  We straightforwardly obtained a solution for arbitrary temperature in the range of interaction strengths from weak to strong coupling ($0< U < 20\Delta$). Here we present results for $U=12 \Delta$ in the entire temperature range. We plotted the temperature dependence of the dimensionless Kondo scale $a$ and of the effective interaction strength $\Lambda$  in Fig.~\ref{fig:A-Lambda}. We can clearly see the saturation of both variables at very low temperatures where the static approximation is best justified ($a \ll 1$). The approximation extended to higher temperatures still delivers qualitatively reliable results. It asymptotically goes over to the Hartree approximation in the high-temperature limit ($a\to 1$ and $\Lambda\to U$).     

\begin{figure}[tbh]
\begin{minipage}{0.5\textwidth}\includegraphics[width=7.5cm]{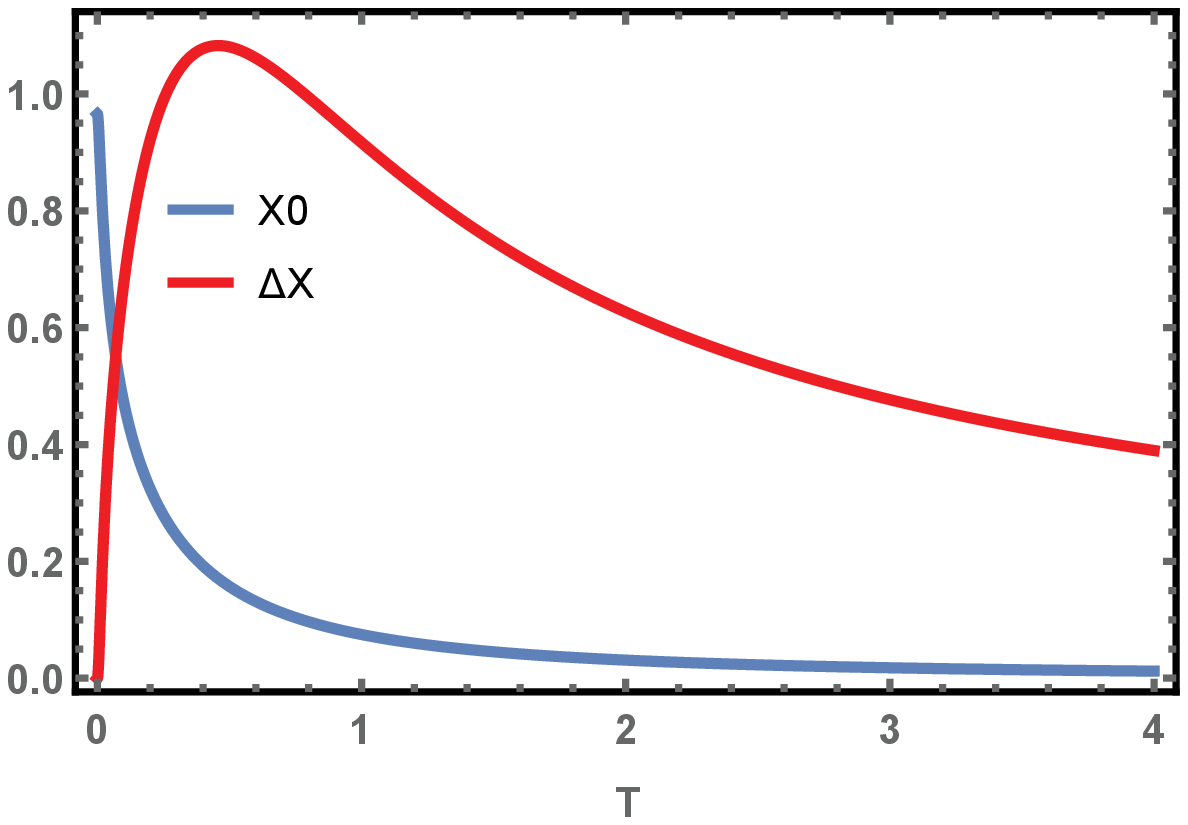}\end{minipage}\begin{minipage}{0.5\textwidth}\includegraphics[width=7.5cm]{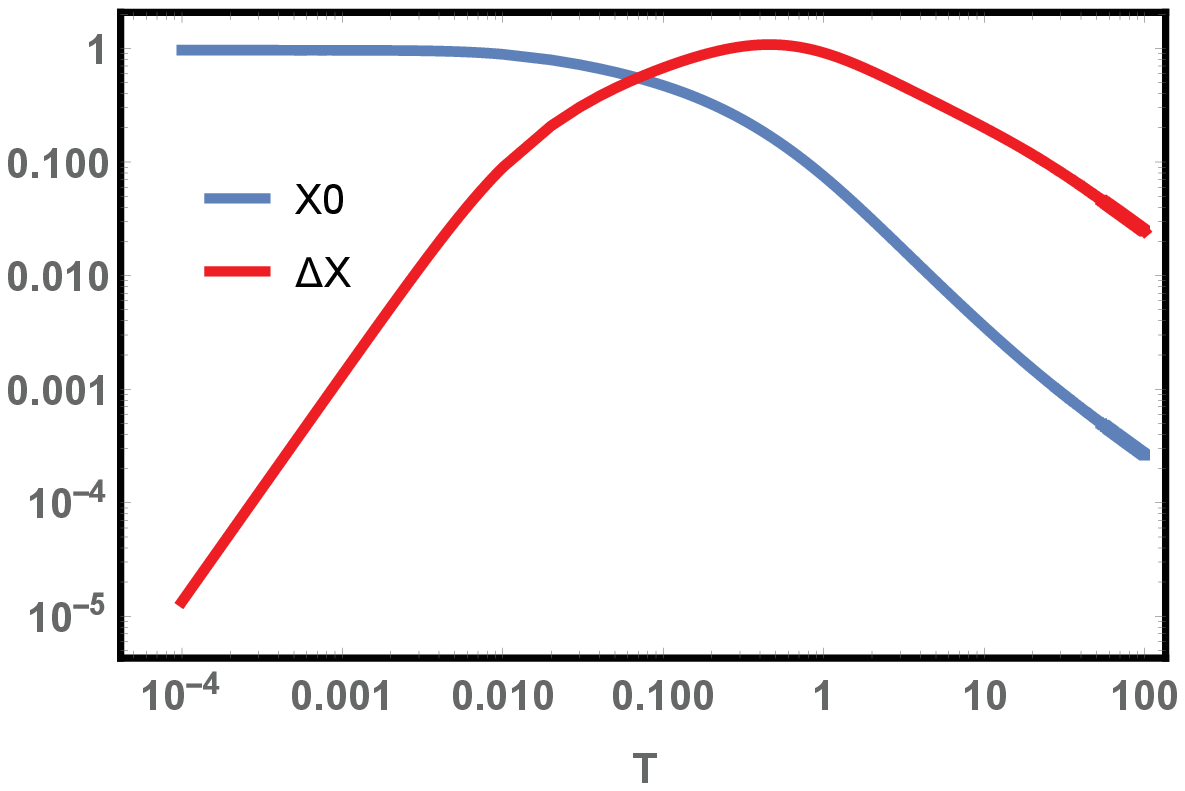}\end{minipage}
\caption{Temperature dependence of thermal ($\Delta X$) and quantum ($X_{0}$) fluctuations contributing to screening the bare interaction and suppressing the spurious singularity of the Hartree approximation. Their crossing determines the Kondo temperature at which the quantum fluctuations overtake control of the low-temperature behavior. Linear scale at left panel and log-log scale at the right panel.}
\label{fig:X0-dX}
\end{figure}

The substantial contributions to the renormalization of the bare interaction come from the $X$ integrals from Eqs.~\eqref{eq:X}. Integral $X_{0}$ monotonically increases with decreasing temperature and reaches its maximum at absolute zero where only quantum fluctuations survive. While the behavior of integral $\Delta X$ is different, see Fig.~\ref{fig:X0-dX}. It first increases with decreasing temperature and reaches its maximum at a non-zero temperature. After that it turns down and vanishes at zero temperature. It hence contains only thermal fluctuations in the low-temperature regime.  At a crossover temperature the two contributions equal, $X_{0}(a) =\Delta X(a) $, and we use this crossing as a definition of the Kondo temperature. It is just the temperature at which the quantum and thermal fluctuations equal. 

The Kondo behavior can also be observed at the temperature dependence of the magnetic susceptibility. Strong electron correlation causes forming of the local magnetic moment in the single impurity Anderson model. However, due to the electron-hole symmetry in the local Fermi liquid, the singlet electron-hole pair decays at time scales inversely proportional to the Kondo scale/temperature. The existence of a long living magnetic moment is reflected in the Curie-Weiss form of the magnetic susceptibility. It goes, however, over at the Kondo temperature to the Pauli susceptibility of the Fermi gas. The magnetic susceptibility in the static approximation has the Hartree form where the bare interaction is replaced by the effective one, 
\begin{align}
\chi &\doteq - \frac {2g}{1 + g\Lambda(a)} \,.
\end{align}
\begin{figure}[tbh]
\begin{minipage}{0.5\textwidth}\includegraphics[width=7.5cm]{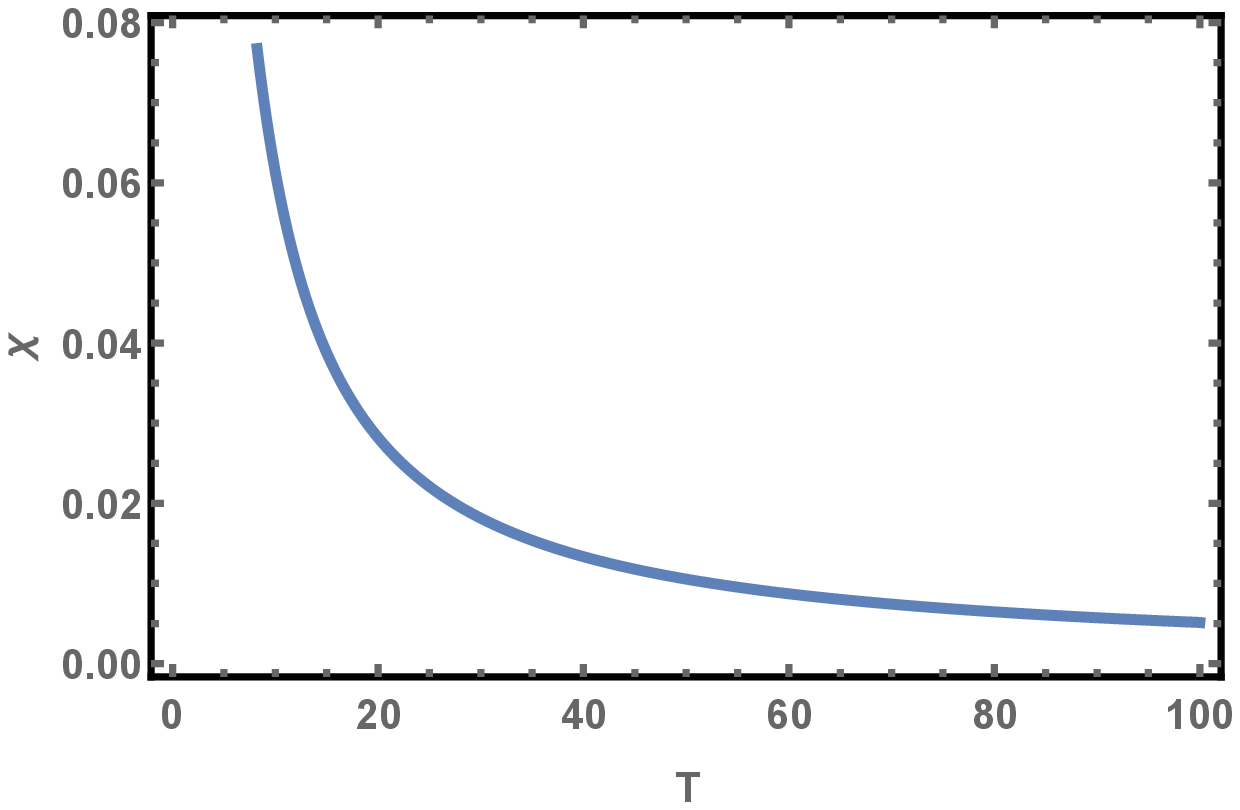}\end{minipage}\begin{minipage}{0.5\textwidth}\includegraphics[width=7.5cm]{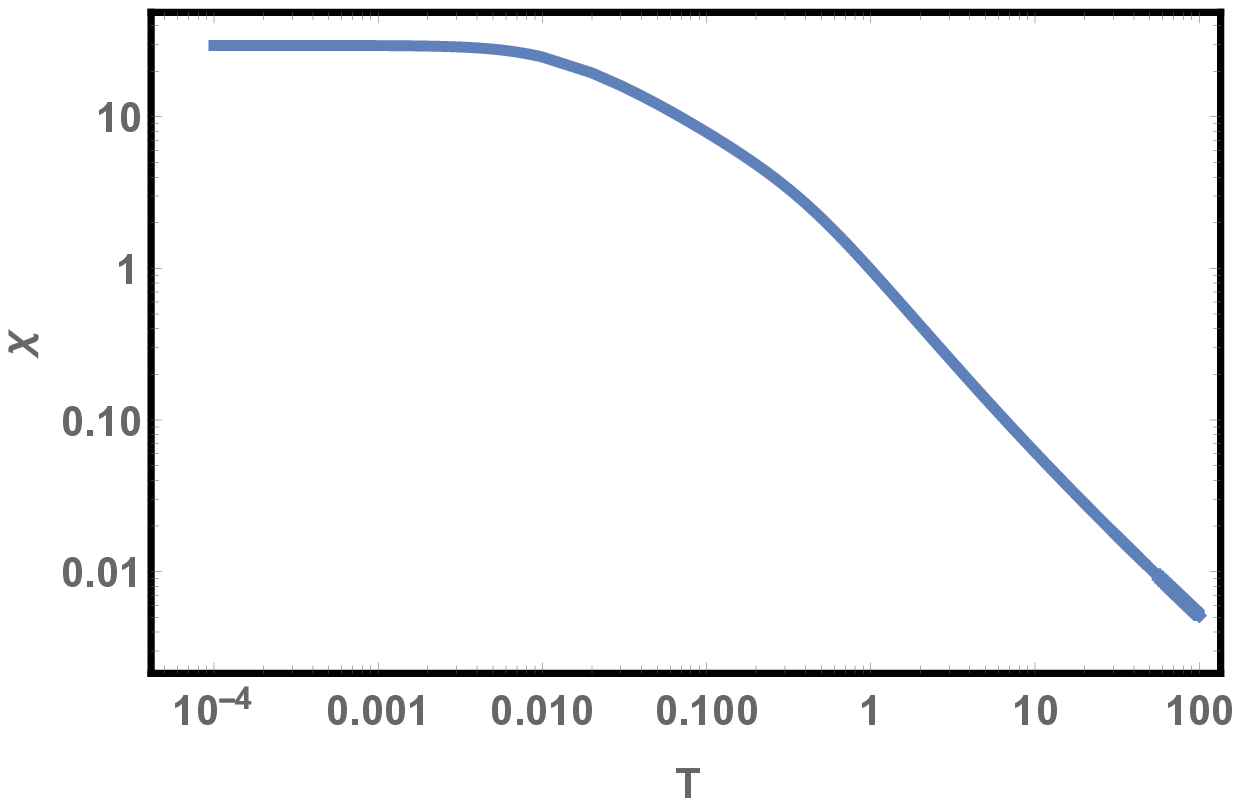}\end{minipage}
\caption{Temperature dependence of the magnetic susceptibility in linear scale (left panel) and log-log scale (right panel). The high-temperature Curie-Weiss $T^{-1}$ behavior saturates at the Kondo temperature and goes over to the Pauli susceptibility when approaching zero temperature.}
\label{fig:chi}
\end{figure}
We plotted its  temperature dependence in Fig.~\ref{fig:chi} where this behavior is clearly demonstrated for $U=12\Delta$.  It essentially behaves as the inverse of the Kondo scale $a$. 

\begin{figure}[tbh]
\begin{minipage}{0.5\textwidth}\includegraphics[width=7.5cm]{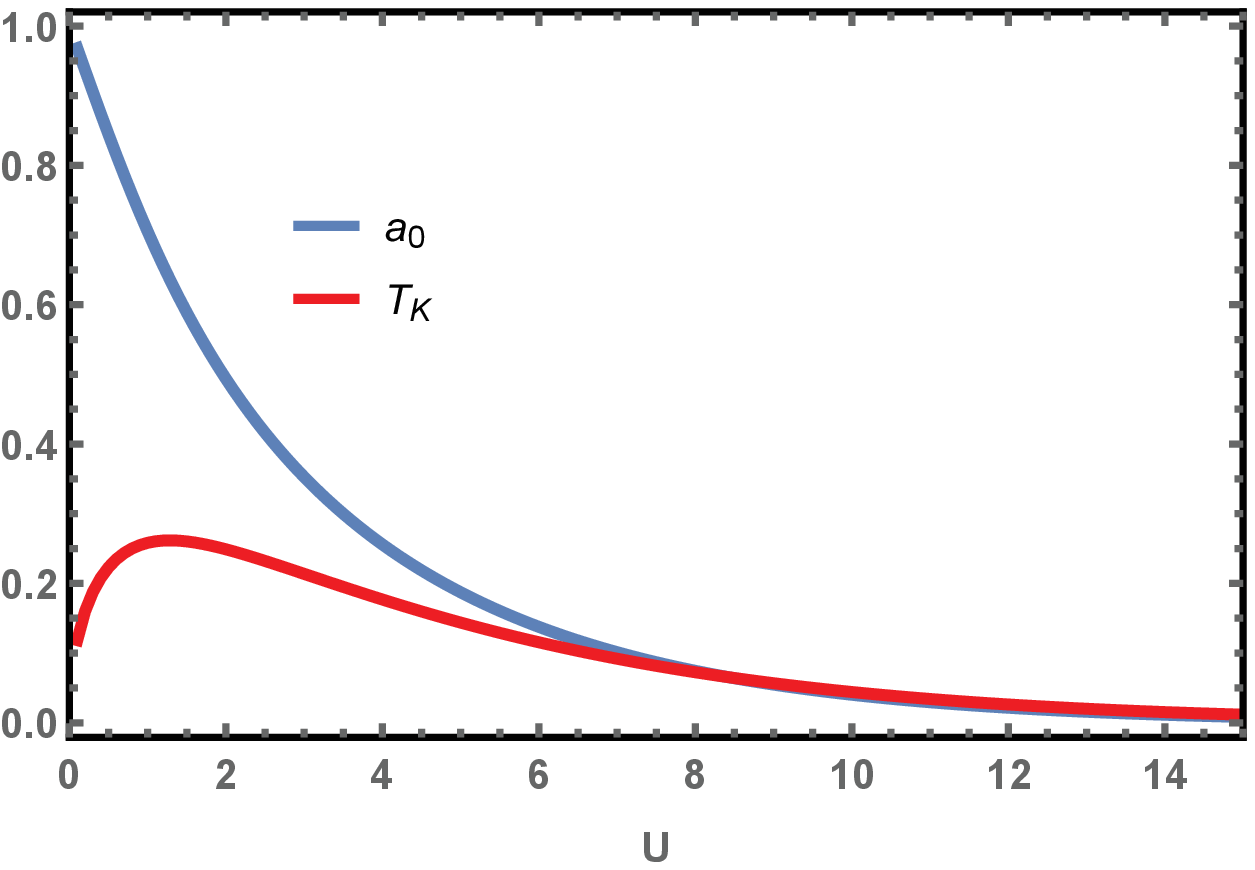}\end{minipage}\begin{minipage}{0.5\textwidth}\includegraphics[width=7.5cm]{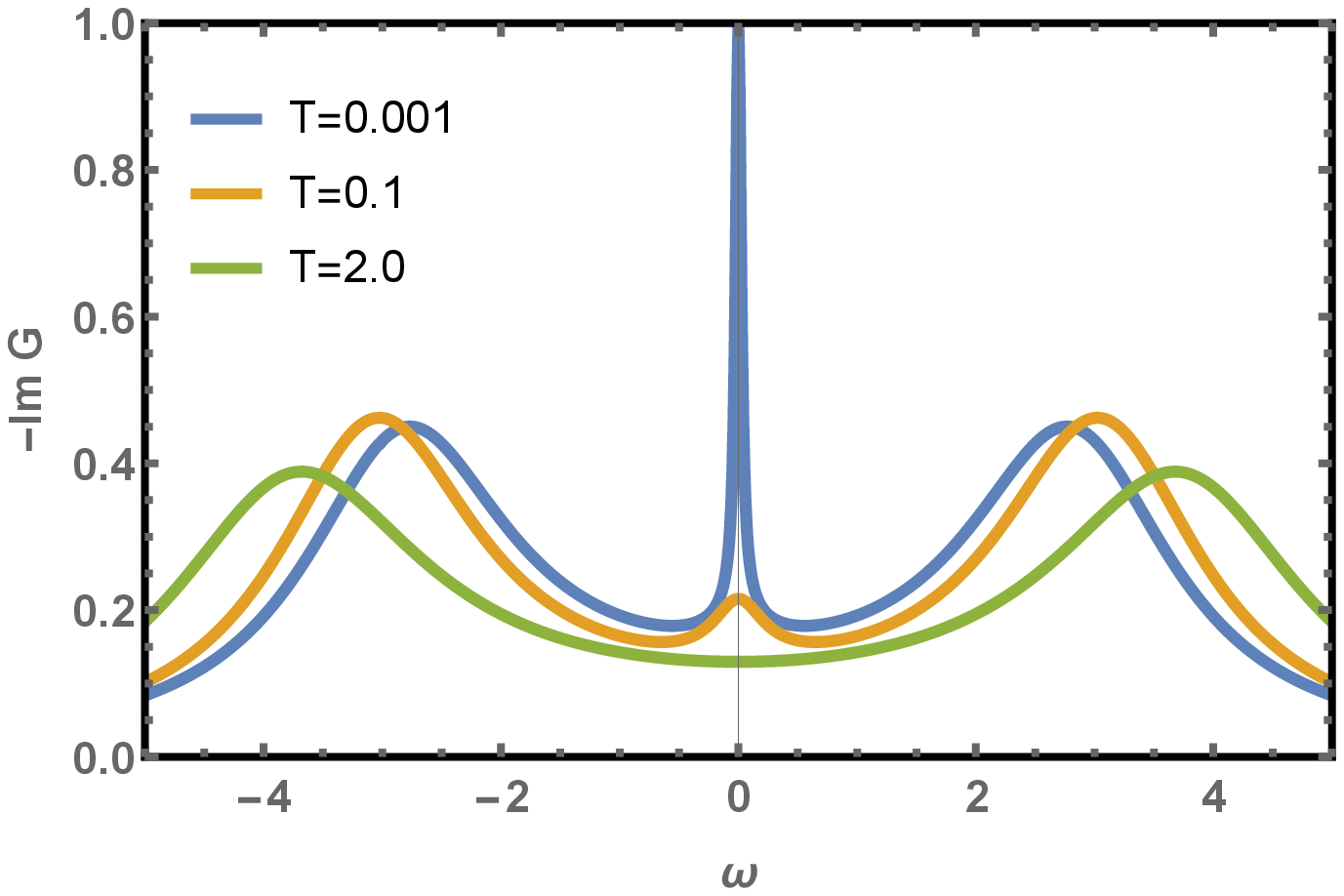}\end{minipage}
\caption{Kondo temperature from the asymptotic expression in Eq.~\eqref{eq:Kondo-Temp} compared to the Kondo scale $a_{0}$ as a function of the interaction strength $U$ (left panel). Typical normalized spectral function in strong coupling, $U=8\Delta$, for a few decreasing temperatures (right panel).}
\label{fig:Kondo}
\end{figure}

Equations~\eqref{eq:X}-\eqref{eq:a-def} have two temperature regimes in the Kondo limit $a\to0$ connected with the contribution to integral $X$. The high-temperature regime is characterized by $\Delta\beta a \ll 1$ with the dominance of $\Delta X$ and Curie-Weiss susceptibility.  While the low-temperature regime is characterized by  $\Delta\beta a \gg 1$  and the Fermi-liquid asymptotics of thermodynamic quantities of order $k_{B}^{2}T^{2}/\Delta^{2}a^{2}$ in integral $X_{0}$.  The two regimes meet at the Kondo temperature. It is exponentially small in the strong-coupling limit and can be well approximated by a formula 
\begin{equation}\label{eq:Kondo-Temp}
\frac{k_{B}^{2}T_{K}^{2}}{\Delta^{2}} = \frac{D_{0}\Delta a_{0}^{2}}{2(D_{0}\Delta - a_{0})^{2}}\left[D_{0}\Delta \ln\left(\frac{D_{0}\Delta}{a_{0}} \right) - D_{0}\Delta + a_{0} \right] \,
\end{equation}  
where subscript $0$ refers to zero-temperature values. The left panel of Fig.~\ref{fig:Kondo} shows the dependence of the Kondo temperature on the interaction strength together with the Kondo scale $a_{0}$. Since Eq.~\eqref{eq:Kondo-Temp} holds only asymptotically for $a_{0}\to 0$, it becomes unreliable in the weak-coupling regime where $a\to 1$ and it becomes meaningless to speak of the Kondo temperature. The agreement between the two scales in strong coupling is excellent. 

\section{Self-energy and spectral function}

Kondo behavior can be demonstrated not only on thermodynamic quantities but also on the one-particle spectral function, where it leads to the formation of a narrow central quasiparticle peak.  We can demonstrate such a behavior if we match the two-particle vertex with the one-particle self-energy. The dynamical self-energy has the same form as in the RPA where the bare interaction is replaced by the effective one, the static irreducible electron-hole vertex $\Lambda_{\uparrow\downarrow}$ \cite{Janis:2019aa}. It is  
\begin{equation}
\Sigma(\omega_{+}) =   U\int_{-\infty}^{\infty}\frac{dx}{\pi } \left\{f(x + \omega) \frac{\mathcal{D}(x_{-}) - 1}{\mathcal{D}(x_{-})}  \Im G(x_{+} + \omega) + b(x)G(x_{+} + \omega) \Im\left[\frac{1}{\mathcal{D}(x_{+})} \right]  \right\} \,.
\end{equation}
%
The full one-electron propagator then is $\mathcal{G}(\omega_{+}) = 1/[\omega + i\Delta - \Sigma(\omega_{+})] $. The genesis of the central quasiparticle peak in the spectral function with decreasing temperature is plotted on the right panel of Fig.~\ref{fig:Kondo}. The temperature fluctuations combined with the Coulomb repulsion generally expel electrons from the Fermi energy, and the spectral function consists of two rather broad peaks separated by a shallow central valley.  The Fermi-liquid starts to form at low temperatures, just around the Kondo temperature, and the Friedel sum rule drives the electrons back to the Fermi energy at which the quasi particle Kondo-Suhl resonance starts to form.

\section{Conclusions}

The Kondo temperature determines a new scale in the Fermi-liquid regime controlling the low-temperature asymptotics of thermodynamic quantities. Although it is generally defined from the zero-temperature limit, we showed that it can also be defined as a crossover temperature at which the thermal fluctuations in the two-particle vertex equal the quantum ones. Such a definition is, however, possible only in theories renormalizing properly the bare interaction as done in an analytic way in the reduced parquet equations of Refs.~\cite{Janis:2017aa}-\cite{Janis:2019aa}. Such a separation of the high (classical) and low-temperature (quantum) fluctuations allows us to use the same concept as well in extended low-dimensional models ($d=1,2$) where the long-range order is prohibited by the Mermin-Wagner theorem  \cite{Mermin:1966aa}.  

\section*{Acknowledgment}
 The research was supported by Grant No. 19-13525S of the Czech Science Foundation. 


\end{document}